\pdfoutput=1 
\documentclass[11pt]{article}
\usepackage[utf8]{inputenc}
\usepackage[DIV12]{typearea}
\usepackage{amsmath,amsfonts,amssymb,amscd}
\usepackage[all]{xy}
\usepackage{graphicx}
\usepackage{cite}
\usepackage{mathrsfs}
\usepackage{bbm}
\usepackage{bbold}
\usepackage{xcolor}
\usepackage[normalem]{ulem}
\usepackage{tensor}
\usepackage{mathtools}
\usepackage[bookmarks=false]{hyperref}
\hypersetup{
    pdftitle = {Symmetries from outer automorphisms},
    pdfauthor = {Bischer,D\"oring,Trautner},
    colorlinks=true,
    linkcolor=black,
    filecolor=black,      
    urlcolor=black,
    citecolor=black
}
\usepackage{srcltx}
\usepackage{tikz}

\newcommand{\Eqref}[1]{Eq.~\eqref{#1}}
\newcommand{\Figref}[1]{Fig.~\ref{#1}}

\newcommand{\SO}[1]{\ensuremath{\mathrm{SO}(#1)}}
\newcommand{\SU}[1]{\ensuremath{\mathrm{SU}(#1)}}

\newcommand{\U}[1]{\ensuremath{\mathrm{U}(#1)}}
\newcommand{\Z}[1]{\ensuremath{\mathbbm{Z}_{#1}}} 

\newcommand{\rep}[2][]{\ensuremath{\boldsymbol{#2}#1}}

\newcommand{\mat}[1]{\left(\begin{matrix}#1\end{matrix}\right)}

\definecolor{darkgreen}{HTML}{109930}

\allowdisplaybreaks[1]

\def\mytitle{Symmetries from outer automorphisms \\[0.2cm] and unorthodox group extensions}

\title{\mytitle}

\begin{document}

\begin{titlepage}
\setlength{\topmargin}{0.0 true in}
{\hfill ULB-TH/24-11\vspace{1cm}}
\thispagestyle{empty}

\vspace*{1.0cm}


\begin{center}
{\LARGE\textbf{\mytitle}}

\renewcommand*{\thefootnote}{\fnsymbol{footnote}}

\vspace{1.6cm}
\large
Christian D\"oring${}^{1,}$\footnote[1]{
\href{mailto:christian.doring@ulb.be}{christian.doring@ulb.be}
}~~and~~
Andreas Trautner${}^{2,}$\footnote[2]{
\href{mailto:trautner@mpi-hd.mpg.de}{trautner@mpi-hd.mpg.de}
}
\normalsize
\\[5mm]
\textit{
${}^1$Service de Physique Th\'eorique, \\ Universit\'e Libre de Bruxelles, C.P. 225, B-1050 Brussels
} \\[5mm]
\textit{
${}^2$Max-Planck-Institut f\"ur Kernphysik \\ Saupfercheckweg 1, 69117 Heidelberg, Germany
}

\end{center}
\vspace*{12mm}

\begin{abstract}\noindent
Symmetries play an essential role in the construction and phenomenology of quantum field theories (QFTs).
We discuss how to construct symmetries of QFTs by extending minimal ``seed'' symmetry groups to larger groups that 
contain the seed(s) as subgroup(s). On the one hand, there are so-called ``normal'' extensions, which are given by outer automorphisms of the original symmetry group (including the trivial one) and contain the seed as a normal subgroup.
On the other hand, there can be ``unorthodox extensions'' which do not have this property. We demonstrate our logic on the most general scalar potentials of the two- and three-Higgs-doublet models (2HDM and 3HDM). For the 2HDM, we show that all symmetry groups, including the different possible classes of CP and continuous symmetry groups, can be obtained from extensions of the smallest possible symmetry CP1 by consecutive outer automorphisms. Scanning over normal and unorthodox group extensions might be the easiest way to ``machine learn'' the possible symmetries of a QFT. However, many of the groups constructible in this way may not be realizable in a concrete model, in the sense that they lead to additional accidental symmetries. Hence, we also comment on a different, ``top-down'' way to obtain the possible realizable symmetry groups of a QFT based on the covariant transformation of couplings under the most general basis changes. 
\end{abstract}
\thispagestyle{empty}
\clearpage

\end{titlepage}

\section{Introduction}
One of the most intensively studied theories beyond the Standard Model (SM) is the 2HDM~\cite{Branco:2011iw}.
This has the pleasant side effect that a lot of details are known about this -- in parts of the parameter space still phenomenologically viable -- quantum field theory~(QFT).
This includes the hard-earned knowledge about \textit{all} of the possible exact\footnote{%
Of course, these symmetries are generally expected to be broken by quantum (incl. quantum gravity) effects and they may also be broken explicitly (incl.\ the possibility of chiral anomalies) once extended to fermions.
However, these effects are of no concern for our study focused entirely on the scalar sector.} 
global symmetries of its most general, renormalizable scalar potential~\cite{Ivanov:2005hg,Ivanov:2006yq,Ferreira:2010yh,Ferreira:2023dke}.
Of interest for phenomenological applications are also approximate symmetries of this potential~\cite{Battye:2011jj,Pilaftsis:2011ed,Dev:2014yca,Darvishi:2019dbh}, and the extension of the symmetries to the SM Yukawa sector, see e.g.~\cite{Ivanov:2013bka,Nishi:2014zla}. 
Not quite as exhaustive studies exist for models with three or more copies of the Higgs field, namely the 3HDM~\cite{Ivanov:2011ae,Ivanov:2012ry,Ivanov:2012fp,Ivanov:2014doa,Fallbacher:2015rea,deMedeirosVarzielas:2019rrp,Darvishi:2021txa,Bree:2024edl}, partly for the 4HDM~\cite{Shao:2023oxt,Shao:2024ibu} and beyond~\cite{Ferreira:2008zy,Ivanov:2010wz,Ivanov:2013bka,Darvishi:2019dbh,Plantey:2024yfm}.

\smallskip
Here, we will not be interested in the 2HDM or NHDM \textit{per se}, but we will use their well-explored symmetry landscape as a laboratory to explore the nature and interplay of symmetries themselves. In particular, we lay out the methodology to extend symmetry groups of any QFT to larger symmetry groups in consistency with the already present symmetries of the theory. Our methods are general and applicable to any QFT.

\smallskip
All symmetries under discussion, including Higgs-flavor as well as general charge-parity conjugation (CP) transformations, can be represented as matrix groups acting linearly on the fields.
Once imposed on the level of the Lagrangian, possible symmetry transformations enforce relations of couplings. 
This makes it easy to fall prey to the misconclusion that the physical consequences of each of the possible matrix groups is somewhat on the same footing.
However, recently one of the authors and collaborators have argued~\cite{Bento:2020jei}
that there are decisive differences between some of the symmetry groups from an algebraic (one might also say topological) point of view, 
which only arise once the effects of the group action on the specific Lagrangian is considered.
In other words, there are some relations of couplings which are inherently different from others.
These inherent differences among the symmetry enforced relations become very apparent once the relations are formulated in terms of basis invariants.\footnote{%
Basis invariants are, by definition, combinations of theory parameters (couplings) that are invariant under field redefinitions or reparametrization of the Lagrangian using all possible freedom of basis choices.}
In the systematic, group-theoretical construction of the invariants~\cite{Trautner:2018ipq,Trautner:2020qqo} one first combines the couplings of a theory
to well-transforming basis covariants~\cite{Ivanov:2019kyh}, which are then contracted to form orthogonal invariants.
does lead to \textit{two different} kinds of possible consequences:
\begin{itemize}
\item[(i)] \textit{Either}, basis covariant objects are forced to obey certain geometrical alignments, corresponding to non-trivial interrelations of orthogonal basis invariants, 
\item[(ii)] \textit{or}, basis covariant objects are forced to vanish, corresponding to vanishing of all basis invariants that contain the corresponding covariant. 
\end{itemize}
The first option leaves the algebraic ring of couplings intact, while the second option ``collapses'' the original ring to a smaller subring~\cite{Bento:2020jei}. 

We firstly point out that this insight reveals the pathway to infer possible (actively acting) symmetries of a model in a ``top-down'' way, starting from subgroups of the (passively acting) group of 
possible basis changes. On the one hand, one searches for transformations that enforce the vanishing of covariants. On the other hand, one searches for transformations
that are preserved once covariants obey specific alignments.
The first problem amounts to formulating linear constraints, while the latter problem is the same as the much more frequently faced challenge of finding all possible breaking patterns of continuous groups down to (possibly discrete) 
subgroups~\cite{Adulpravitchai:2009kd,Luhn:2011ip,Merle:2011vy,Fallbacher:2015pga} with a given set of covariants. Viewing the problem in this way, it also becomes clear why removing one or several of the covariants 
poses a ``topologically'' different question than merely changing their alignment.

The ``top-down'' approach to explore symmetries of a model offers a superior starting point to find all possible realizable\footnote{%
We mean ``realizable'' in the technical sense of~\cite{Ivanov:2010wz}. A realizable symmetry group is one which does not automatically cause the conservation of additional transformations.
}
symmetry groups of a model without having to resort to test the Lagrangian for accidental symmetries. However, we will not follow this route here. Instead, the purpose of the present paper is to show how all possible symmetry groups of a given theory can be constructively obtained from smaller symmetry groups of a model by extending them in a ``bottom-up'' way that we detail below. For the 2HDM and 3HDM we will explicitly show that all such extensions can be found by using \textit{outer automorphisms} of smaller symmetry groups, which are ``group extensions'' in the mathematical sense. This does not hold in general (in particular, if the group that one would arrive after the extension is a simple group) and we will discuss alternative ``unorthodox'' extensions which complete all possibilities for group extensions.

In the following section we discuss general group extensions and their relations to outer automorphisms. Then we turn to the 2HDM as an example. 
We briefly review the 2HDM symmetry groups, the covariant coupling tensors of the potential on which the symmetries act, and the associated symmetry map in section~\ref{sec:2HDMsymmetries}.
Then we show that all exact global symmetry groups of the 2HDM are contained in the outer automorphisms (and outer automorphisms of extensions) of the smallest possible symmetry of the model, CP1. 
Finally, we show that the analogous procedure works for the 3HDM and comment other models where the inclusion of unorthodox extensions can become necessary. 

\section{Extensions of symmetry groups}\label{sec:extensions}
\subsection{Generalities}\label{sec:extensions_generalities}
Take a QFT containing a generic vector of quantum fields $\vec\Phi(x)$ that is invariant under the linear action of some symmetry group $G$, mapping $\vec\Phi\mapsto \mathcal{X}(\mathsf{g})\vec\Phi$, where 
$\mathsf{g}\in G$ is an element of $G$ and $\mathcal{X}(\mathsf{g})$ denotes the matrix representation of $\vec\Phi$ under $G$.
We are interested in \textit{extensions} of the symmetry group $G$, in such a way that the original theory obtains a bigger symmetry group $\Gamma$, which contains $G$ as a subgroup, $G\subset\Gamma$. 
Assuming we want to extend $G$ by a new generator $\mathsf{h}$, the bigger group $\Gamma$ can always be obtained as closure of $\mathsf{h}$ and $G$, i.e.~$\Gamma=\langle\mathsf{h},G\rangle$.\footnote{%
For our physically intuitive access here, we will not be bothered by spelling out all mathematical subtleties in detail. Whenever in doubt, one should retract to explicit matrix representations of all groups.}
There are two mutually exclusive possibilities for such an extension, namely
\begin{align}\label{eq:group_extension}
(i):\qquad \mathsf{h}\,G\,\mathsf{h}^{-1}~&=~G\; \qquad\qquad\text{(normal group extension),}~\quad\text{or}~\\ \label{eq:unorthodox_extension}
(ii):\qquad \mathsf{h}\,G\,\mathsf{h}^{-1}~&=~G'~\neq~G\; \quad\text{(unorthodox extension).}
\end{align}
Case $(i)$ has been studied in great detail and is known as a ``group extension'' in the mathematical literature, see~\cite{Ivanov:2012ry,Ivanov:2012fp} for a physicist friendly account.
The second case $(ii)$ appears to be not so well studied and we term it an ``unorthodox'' extension.

\begin{figure}[!t!]
\includegraphics[width=0.30\linewidth]{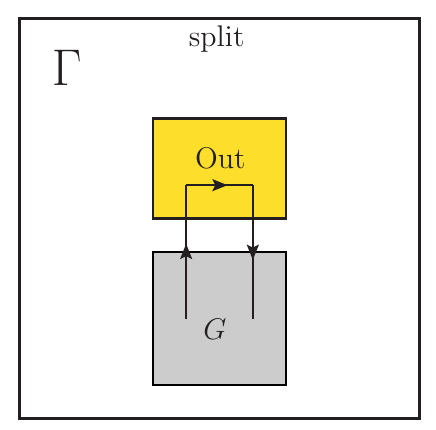}%
\hfill
\includegraphics[width=0.30\linewidth]{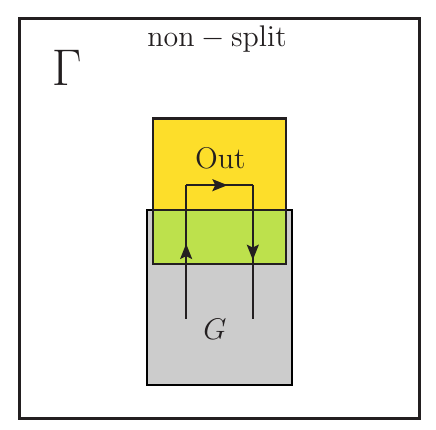}%
\hfill
\includegraphics[width=0.30\linewidth]{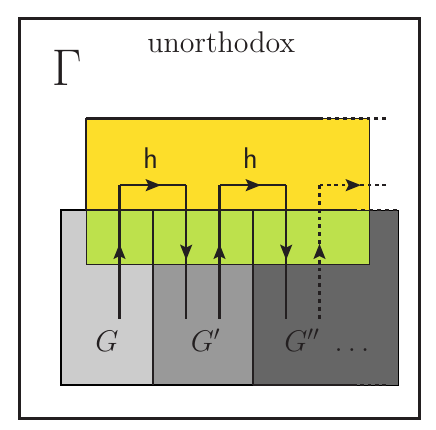}%
\caption{\label{fig:extensions}%
Cartoon of split and non-split normal group extensions of a group $G$ by an outer automorphism $\mathrm{Out}(G)$ to a larger group $\Gamma$ (left and middle panel). We also show unorthodox extensions by a new generator $\mathsf{h}$ (right panel). Unorthodox extensions can also be either split~(not shown) or non-split, and they can be finite or infinite.}
\end{figure}

In the first case, $(i)$, $G$ is a normal subgroup of $\Gamma$ denoted as $G\triangleleft\Gamma$, and $\mathsf{h}$ has to act as an automorphism of $G$. 
It is possible that some (non-trivial) elements of the group $H:=\langle\mathsf{h}\rangle$ are already contained in $G$, which is called a non-split extension; in this case $\Gamma/G\neq H$.\footnote{%
Our definition of $H$ here, which is a group, should not be confused with the more common definition of $\mathcal{H}$ as complement of a normal subgroup $\mathcal{N}$ inside a bigger group $\mathcal{G}$, in which case $\mathcal{H}$ would be a coset and $\mathcal{G}/\mathcal{N}=\mathcal{H}$ holds by definition.}
Alternatively, $\Gamma/G=H$ which is called a split extension implying that $\Gamma$ can be constructed as a semi-direct product $\Gamma=G\rtimes H$.
The split and non-split extensions are schematically depicted in the left two panels of Fig.\,\ref{fig:extensions}.

In the second case, $(ii)$, $G$ is not normal in $\Gamma$. Instead, the action of $\mathsf{h}$ maps $G\subset\Gamma$ to a conjugate subgroup $G'\subset\Gamma$, which still ensures that $G\cong G'$ are isomorphic. 
In a sense, $\mathsf{h}$ generates a translation from one conjugate subgroup to another. Depending on whether parts of $H$ are contained in these subgroups, one may also distinguish split and non-split unorthodox extensions.
Investigating the mathematical details of such constructions may be interesting in view of the general problem of group extensions, but is beyond the scope of this work. A schematic depiction of 
a (non-split) unorthodox extension is shown in the right panel of Fig.\,\ref{fig:extensions}.

To be very clear, we remark that it is, in general, possible to extend the symmetries of a model by \textit{any} linearly acting transformation $\vec\Phi\mapsto \mathcal{M}\vec\Phi$. 
In our language, $\mathcal{M}=\mathcal{M}(\mathsf{h})$ represents the generator $\mathsf{h}$ of the new symmetry group $H$, and the total group is simply given by the closure of $G$ and $H$, which is commonly denoted as $G\cup H$.
Explicitly, the closure can be constructed from all possible matrix power products (monomials) of $\mathcal{M}(\mathsf{h})$ and $\mathcal{X}(\mathsf{g})$, $\forall~\mathsf{h}\in H$ and $\forall~\mathsf{g}\in G$.

The possibilities of normal and non-normal subgroup of $G\subset\Gamma$ are mutually exclusive and exhaustive. This implies that the above options of normal and unorthodox extensions cover \textit{all} possible symmetry extensions.

\subsection{Symmetry extensions from Outer Automorphisms\label{sec:SymfromOuterAuts}}
Let us focus on normal group extensions. From Eq.~\eqref{eq:group_extension} it is clear that $\mathsf{h}$ must generate an automorphism of $G$. 
Since inner automorphisms are already contained in $G$ they do not lead to new additional symmetries. Hence, we can focus on \textit{outer} automorphisms.
Outer automorphisms (Outs) are, literally speaking, symmetries of symmetries~\cite{Trautner:2016ezn,Trautner:2017vlz}.
For the present work it will largely be sufficient to view Outs as \textit{redundancies} of a symmetry group. 
That is, in a sloppy way we might say that we are looking for transformations that are consistent with (i.e.\ leave invariant) 
the set of all already present symmetry transformations (leaving invariant the set does not necessarily mean leaving invariant each element in the set).\footnote{%
The fact that the full set of symmetry transformations is left invariant implies that also the functional form of the potential is left invariant. Hence, the action of Outs will only act as a linear transformation on the couplings of the potential, for details see~\cite{Fallbacher:2015rea}, and, therefore, also imply a well-defined transformation on the 
basis-covariant combination of the couplings.}
We spare the abstract group theory for most parts and focus on explicit representations.
Nonetheless, we start by stating the most general consistency condition for Outs to show how it reduces to our case.

In an abstract sense, the Outs $u$ of a group $G$ act as permutation of the group elements \mbox{$u:\,\mathsf{g}\mapsto \mathsf{g'}$} for all $\mathsf{g},\mathsf{g'}\in G$.\footnote{%
Despite the fact that Outs are strictly speaking cosets of automorphisms (defined only up to an inner automorphism) it helps to 
pick a specific representative of an inner automorphism and imagine $u(\mathsf{g})$ very concretely as a specific permutation of the group elements.}
In a concrete sense, this implies that Outs act as specific permutations of the irreducible representations (irreps) of a group, 
\mbox{$\mathrm{Out}:\,\rep{r}\mapsto\rep{r'}$} where \rep{r} and \rep{r'} denote irreps of $G$. 
In order to be consistent with the group structure, the corresponding transformation matrix $U$ 
of an Out has to fulfill the consistency condition~\cite{Fallbacher:2015rea, Trautner:2016ezn} 
\begin{equation}\label{eq:ConsistencyCondition}
 U\,\rho_{\rep{r'}}(\mathsf{g})\,U^{-1}~=~\rho_{\rep{r}}(u(\mathsf{g}))\;,\qquad\forall \mathsf{g}\in G\;,
\end{equation}
where $\rho_{\rep{r}}(\mathsf{g})$ denotes the irreducible matrix representation of a group element $\mathsf{g}$ in irrep $\rep{r}$
and $u(\mathsf{g})$ is the outer automorphism transformation of the abstract group elements. In particular, this may include the special cases $\rep{r'}=\rep{r}$ or $\rep{r'}=\rep{r}^*$, and also the special case of a trivial outer automorphism $u(\mathsf{g})=\mathsf{g}$. 
After having made a basis choice for \rep{r} and \rep{r'}, $U$ is determined by the consistency condition, Eq.~\eqref{eq:ConsistencyCondition}, up to a 
central element of the group and a global phase. Note that on the level of reducible representations there may be additional redundancies in the action of the Out~\cite{Bischer:2022rvf} and we also exploit this in the present work. 

\section{Symmetries of the 2HDM}\label{sec:2HDMsymmetries}
We now discuss the possible exact global symmetry groups of the 2HDM scalar potential as a concrete example of our abstract considerations above. 
\subsection{Multi-Higgs scalar potential}
The four real degrees of freedom of each Higgs field are arranged as two complex scalar fields $\Phi_a(x)\equiv (\phi^+_a(x)\,,\,\phi^0_a(x))^{\mathrm{T}}$ such that $\Phi_a(x)$ transforms in a representation $(\rep{2},-1/2)$ under the SM gauge group $\SU{2}_\mathrm{L}\times\U1_\mathrm{Y}$. The full scalar potential of an NHDM with $a,b,c,d=1,\dots,N$ then can be written as (summation over repeated indices is implied)
\begin{equation}\label{eq:Potential}
 V=\tensor{\hat Y}{^a_b}\left(\Phi_a^\dagger\,\Phi^b\right)+\tensor{\hat Z}{^a^b_c_d}\left(\Phi_a^\dagger\,\Phi^c\right)\left(\Phi_b^\dagger\,\Phi^d\right)\;. 
\end{equation}
Hermiticity and gauge invariance put constraints on the coupling tensors $\hat{Y}$ and $\hat{Z}$, see e.g.~\cite{Botella:1994cs,Gunion:2005ja,Trautner:2018ipq}.
For the 2HDM, $a=1,2$ and we collect the identical copies of Higgs fields in a two dimensional vector $\Phi(x)\equiv\left(\Phi_1(x),\Phi_2(x)\right)$.
For the 2HDM, there are $14$ real degrees of freedom (real coupling coefficients) in $\hat{Y}$ and $\hat{Z}$. As we discuss in detail below, the freedom of Higgs basis changes can further be used to absorb three 
of the $14$ parameters, resulting in the commonly stated $11$ physical parameters of the 2HDM scalar sector.

\subsection{2HDM symmetries}
The global symmetry groups of the 2HDM scalar potential can be arranged in the sequence~\cite{Ferreira:2009wh,Ferreira:2010yh,Branco:2011iw}
\begin{equation}\label{eq:symtree}
 \mathrm{CP1}~<~\Z2~<~\left\{\begin{array}{cc} \U1 \\ \mathrm{CP2} \end{array}\right\}~<~\mathrm{CP3}~<~\SU2\;,
\end{equation}
where ``\,$G< \Gamma$\,'' here means that the symmetry of class $G$ of the scalar potential is implied by the symmetry of class $\Gamma$ (but not necessarily a subgroup).

To discuss Higgs-flavor as well as general CP transformations on the same footing, 
we stress that all of the possible 2HDM symmetries can be represented as linearly acting matrix groups. To show this explicitly, we combine the Higgs fields and their 
respective CP conjugates into a single vector 
\begin{equation}
\vec{\Phi}:=\left(\Phi(t,\vec{x}),\Phi^*(t,-\vec{x})\right)^\mathrm{T}\;.
\end{equation}  
Working in this reducible $\rep{r}\oplus\rep{r}^*$ space has the advantage that 
all transformations act linearly on $\vec\Phi$ such that we can seamlessly combine flavor-type and general CP transformations by plain matrix multiplication~\cite{Holthausen:2012dk}.

We will from now on suppress the spacetime arguments as they are irrelevant for the present discussion.
The flavor- and general CP-type transformations therefore act as
\begin{align}
    &\mat{\Phi^{\hphantom{*}}\\ \Phi^*}\mapsto\mat{  S\,&  0\\   0&  S^*\,}\mat{\Phi^{\hphantom{*}}\\ \Phi^*} \quad \text{(Higgs-flavor),} \quad \text{or} \quad \\
    &\mat{\Phi^{\hphantom{*}}\\ \Phi^*}\mapsto\mat{  0&  X\\   X^*&  0}\mat{\Phi^{\hphantom{*}}\\ \Phi^*} \quad \text{(General CP),}
\end{align}
where $S$ and $X$ are unitary {$2\times 2$} matrices.

\smallskip\noindent
For the Higgs-flavor symmetries listed in Eq.~\eqref{eq:symtree}, 
possible choices for the matrices are
\begin{itemize}
    \item $ \mathbbm{Z}_2:\quad\quad\;\,  S=\mat{1&0\\0&-1}$,
    \item $ \text{U(1)}:\quad\;\;   S=\mat{e^{-i\xi}&0\\0&e^{i\xi} }$,
    \item $ \text{SU(2)}:\quad   S=\mat{e^{-i\xi}\cos\theta&-e^{-i\psi}\sin\theta\\ e^{i\psi}\sin\theta&e^{i\xi}\cos\theta }$.
\end{itemize}
The general CP symmetries can be represented by
\begin{itemize}
    \item $ \text{CP}1:\quad   X=\mathbbm{1}_2$,
    \item $ \text{CP}2:\quad   X=\mat{0&-1\\1&0}~\equiv~\varepsilon$,
    \item $ \text{CP}3:\quad   X=\mat{\cos\theta&-\sin\theta\\\sin\theta&\phantom{-}\cos\theta }$.
\end{itemize}
Here $\xi$, $\psi$, and $\theta$ are group parameters that are allowed to take arbitrary real values.

All these matrices are, of course, basis dependent, i.e.\ they generally take different forms in different bases.
Insisting that they should have the above form selects a fixed basis, sometimes up to some residual freedom.
It is straightforward to show that for a basis rotation $\Phi'=U\Phi$, with $U\in\U{2}\cong\U{1}\otimes\SU{2}$ (which should not be confused with the gauge group), the corresponding basis-rotated matrices are given by
\begin{align}
S'=USU^\dagger\;,\qquad \text{and}\qquad X'=UXU^{\mathrm{T}}\;.
\end{align}
The global \U1 factor of the basis change is irrelevant, as it can always be removed by a $\U1_\mathrm{Y}$ hypercharge gauge transformation. 
This implies that we are, in principle, only interested in the projective part of the respective groups and their representations.

\subsection{Basis invariant 2HDM symmetry map}
Conversely to the symmetry transformation matrices, a general basis rotation also affects the couplings in the scalar potential. 
While $\Phi$ transforms in the fundamental, the coupling tensors transform as the direct sum of their irreducible covariant building blocks, as we will discuss next.

It is very useful to decompose $\hat{Y}$ and $\hat{Z}$ into components that transform covariantly under unitary basis changes in Higgs-flavor space,
i.e.\ redefinitions of the type $\Phi'=U\Phi$, with $U\in\U{1}\otimes\SU{2}$. The coupling tensors decompose into irreducible representations of the global 
$\SU{2}$ as (see \cite{Trautner:2018ipq} for more details)
\begin{equation}\label{eq:decomp}
 \hat{Y}~\hat{=}~\rep{1}\,(Y_{\rep{1}})~\oplus~\rep{3}\,(Y)\;,\qquad\text{and}\qquad \hat{Z}~\hat{=}~\rep{1}\,(Z_{\rep{1}_1})\oplus\rep{1}\,(Z_{\rep{1}_2})\oplus\rep{3}\,(T)\oplus\rep{5}\,(Q)\;.
\end{equation}
There are, in general, $14$ real degrees of freedom (real coupling coefficients) in $\hat{Y}$ and $\hat{Z}$,
and in the brackets we have defined our names for the respective irreducible objects that host these degrees of freedom. 
There are three linear combinations of potential parameters, $Y_{\rep{1}}$, $Z_{\rep{1}_1}$, and $Z_{\rep{1}_2}$, which are basis invariant by themselves.
Furthermore, there is a quintuplet $Q$, and the two triplets $T$ and $Y$ which are also linear in the couplings but transform covariantly under basis transformations. 
These constitute the so-called building blocks from which non-linear basis invariants are constructed by mutual contraction. 
One can use the \SU{2} freedom of basis changes to absorb three of the $14$ parameters, resulting in $11$ physical
parameters. These correspond to the $11$ algebraically independent basis invariants 
that are given by all eight possible independent contractions of $Q$, $Y$, and $T$, next to the three linear basis invariants $Y_{\rep{1}}$, $Z_{\rep{1}_1}$, $Z_{\rep{1}_2}$. 

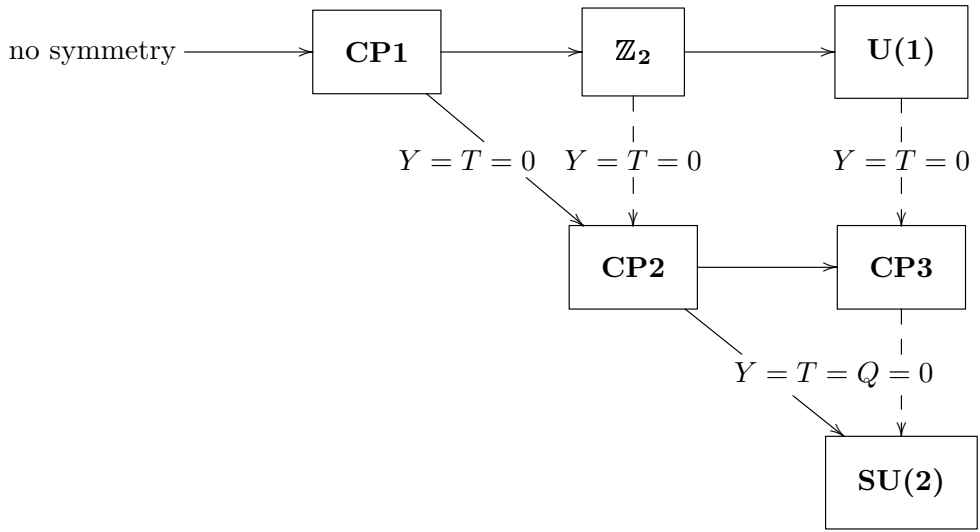
\begin{figure}[t]\renewcommand{\arraystretch}{1.0}
\hspace{\fill}%
 \xymatrix@R+2pc@C+2pc{
 \text{no symmetry} \ar[r] &
 *++++[F]{\txt{\textbf{CP1} } } 
          \ar[r] \ar@{->}[dr]|*+{\hspace{-1cm}Y=T=0}  &  
 *++++[F]{\txt{$\boldsymbol{\mathbbm{Z}_{2}}$ } }   \ar[r] \ar@{-->}[d]|*+{Y=T=0} & 
 *++++[F]{\txt{\textbf{U(1)}  }} \ar@{-->}[d]|*+{Y=T=0} \\ 
 &
 &
 *++++[F]{\txt{\textbf{CP2}}} \ar[r] \ar@{->}[dr]|*+{\hspace{1.75cm}Y=T=Q=0} &  
 *++++[F]{\txt{\textbf{CP3}}} \ar@{-->}[d]|*+{\phantom{Y=T=Q=0}} \\
 &
 & 
 &
 *++++[F]{\textbf{SU(2)}} \\
}
\hspace{\fill}
 \caption{\label{fig:SymmetryMap}
 The symmetry map of the 2HDM scalar sector, adapted from~\cite[Fig.~1]{Bento:2020jei}.
 Horizontal steps correspond to enforcing non-trivial relations among the existing basis invariants of a ring,
 which is one-to-one with enforcing specific alignments of covariants.
 Vertical steps correspond to entirely removing the indicated covariant objects,
 which coincides with collapsing the ring of invariants to a smaller subring.
 Here we show that symmetry enhancements along all arrows are given by extensions of smaller symmetry groups.
 Solid arrows correspond to normal extensions by outer automorphisms (which are all trivial outer automorphisms in this case).
 Dashed arrows correspond to unorthodox extensions.
 }
\end{figure}

Together, the complete set of basis invariants forms an algebraic ring that has firstly been constructed in~\cite{Trautner:2018ipq}.
Following this, necessary and sufficient conditions for the global symmetries in the form of relations among the basis invariants 
have been derived in~\cite{Bento:2020jei}. In \cite{Bento:2020jei} it was also observed that the 2HDM symmetries can, as an improvement to the sequence shown in \eqref{eq:symtree}, 
be very well arranged in the form of a two-dimensional ``symmetry map'', see \Figref{fig:SymmetryMap}. 
All horizontal steps in this symmetry map (solid arrows) are given by relating different, a priori independent, basis invariants (see \cite[Fig.~1]{Bento:2020jei} for the detailed relations).
By contrast, all the vertical steps (dashed arrows) are given by eliminating individual covariant building blocks from the ring, i.e.\ enforcing them to be zero. 
In this sense, each horizontal line represents a ``strand'' of symmetries of an ``intact'' ring in which no degeneracies arise,
while moving vertically requires to ``collapse'' the ring to a smaller (sub-)ring by eliminating building blocks. 
There is a simple geometric picture for this (which works at least for cases known to us): Each horizontal step correspond to a geometric alignment of the various 
non-vanishing building blocks, and the number of preserved symmetry transformations increases from left to right by having an increasing number of (mutually consistent) geometrical alignment conditions. 
By contrast, a vertical step corresponds to entirely removing geometrical objects of a certain kind.
The relative geometric alignments are particularly intuitive in the case of the 2HDM discussed here, because locally $\SU2\cong\SO{3}$ 
and we are only dealing with $\rep{3}$-plets (vectors of $\SO{3}$) and a $\rep{5}$-plet (real symmetric and traceless $3\times3$ matrix).
The corresponding geometrical alignments of these objects and the respective classes of symmetries have been discussed in~\cite{Ferreira:2010yh}.
Such a geometric intuition may not necessarily exist for larger representations and/or larger groups, but we stress 
that it is also not required to exist for the general methodology employed in~\cite{Bento:2020jei}.

We will now proceed to derive the symmetries of the 2HDM from redundancies of the CP1 transformation. We remark that the problem of finding the possible global symmetries of the potential can also be rephrased as a
different, but closely related challenge frequently faced in model building, see e.g.~\cite{Adulpravitchai:2009kd,Merle:2011vy,Luhn:2011ip,Fallbacher:2015pga}: Namely, ``what are the possible subgroups attainable in a breaking of
\SU{2} by representations $\rep{3}$, $\rep{3}$ and $\rep{5}$? And what are the corresponding necessary alignments of the covariants to break to these subgroups?'' In this way it is also very clear why removing one or several of the building blocks poses a 
``topologically'' different question than merely changing their alignment.

\section{Symmetries of the 2HDM from outer automorphisms\label{sec:2HDMSymfromOuterAuts}}
\subsection{Flavor-type and CP-type outer automorphisms\label{sec:TypesOfOuts}}

The matrix groups of explicit symmetry transformations act on the Higgs fields as
\begin{equation}
 \mat{\Phi^{\hphantom{*}}\\ \Phi^*}\mapsto \mathcal{X}(\mathsf{g}) \mat{\Phi^{\hphantom{*}}\\ \Phi^*}\;, \qquad\forall \mathsf{g}\in G\;,
\end{equation}
where we denote the explicit representation matrices by $\mathcal{X}(\mathsf{g})$.
We will focus on normal group extensions here, see Sec.~\ref{sec:extensions}, and ignore for now the possibility of unorthodox extensions.
Hence, any additional transformation with transformation matrix $\mathcal{M}$ and action $\vec\Phi\mapsto \mathcal{M}\vec\Phi$ has to be consistent with the transformations 
in $G$ in the sense that it has to solve Eq.~\eqref{eq:ConsistencyCondition}, 
which here takes the form
\begin{equation}\label{eq:CChere}
 \mathcal{M}\,\mathcal{X}(\mathsf{g})\,\mathcal{M}^\dagger\stackrel{!}{=}\mathcal{X}(u(\mathsf{g}))\;,\qquad\forall \mathsf{g}\in G\;,
\end{equation}
for some automorphism $u$ of $G$.

There are additional requirements on $\mathcal{M}$ once we consider its action $\vec\Phi\mapsto \mathcal{M}\vec\Phi$:
\begin{enumerate}
    \item 
    $\mathcal{M}$ should be unitary to leave the kinetic terms invariant,
    \begin{equation}
     \mathcal{M}^{\dagger}\mathcal{M}\stackrel{!}{=}\mathbbm{1}_4\;.
    \end{equation}
    \item 
    In order to be consistent with the $\SU2_\mathrm{L}\otimes\U1_{\mathrm{Y}}$ gauge symmetry, $\mathcal{M}$ must split into blocks as either
    \begin{equation}
    \mathcal{M}_{\mathrm{flav}}=\mat{  A&  0\\  0&  D}\;,\quad\text{or}\quad
    \mathcal{M}_{\mathrm{CP}}=\mat{  0&  B\\  C&  0}\;.
    \end{equation}
    That is, also the representation of the outer automorphism here has to be either a flavor- or a general CP-type transformation, respectively.
    \item
    Finally, for internal consistency of a symmetry transformation acting on $\vec\Phi=(\Phi,\Phi^*)^\mathrm{T}$
    one should require that 
    \begin{equation}
    C=B^*\;,\quad\text{and}\quad D=A^*\;.  
    \end{equation}
    Any other choice here would correspond to choosing a relative basis between $\Phi$ and $\Phi^*$ which we want to avoid, 
    again with reference to canonical kinetic terms.    
\end{enumerate}
\smallskip
This leaves us with the most general possible forms of $\mathcal{M}$ given by
\begin{equation}\label{eq:generalM}
    \text{flavor-type:}~\mathcal{M}_{\mathrm{flav}}=\mat{  A&  0\\  0&  A^*}\;,\qquad\text{or}\qquad
    \text{CP-type:}~\mathcal{M}_{\mathrm{CP}}=\mat{  0&  B\\  B^*&  0}\;,
    \end{equation}
with unitary matrices $A$ and $B$.

\subsection{Types of Outs of CP1}
We now consider outer automorphisms of the smallest possible symmetry CP1, and show that these contain the generators necessary to generate all classes of symmetries of the 2HDM besides \SU{2}.
This shows that all symmetry groups besides \SU{2} can be derived as normal group extensions of CP1. We will also show that \SU{2} can be derived as unorthodox extension of CP1, or as normal extension of CP2.

For CP1, $\mathcal{X}$ can be chosen to take the form
\begin{equation}\label{eq:CP1}
 \mat{\Phi^{\hphantom{*}}\\ \Phi^*}\xmapsto{\mathrm{CP1}} \mathcal{X} \mat{\Phi^{\hphantom{*}}\\ \Phi^*} = \mat{  0&  \mathbbm{1}_2\\   \mathbbm{1}_2&  0}\mat{\Phi^{\hphantom{*}}\\ \Phi^*}\;.
\end{equation}
As a matrix group, $\mathcal{X}$ generates the two-element group\footnote{%
This \Z2 should not be confused with the \Z2 class of symmetries of the 2HDM referred to in \Eqref{eq:symtree}.} \Z2 which has none but the trivial outer automorphism.
While the Out acts trivially on the level of the abstract group, it may have a non-trivial behavior on the level of the 
representations.\footnote{%
That a trivial automorphism may have a non-trivial action on representations is nothing unusual. For example, it occurs for all elements in the center of a group.}
Hence, to find the possible non-trivial matrix representation of the trivial outer automorphisms we solve the corresponding consistency condition for a general matrix $\mathcal{M}$,
\begin{equation}\label{eq:ccCP1}
 \mathcal{M}\,\mathcal{X}\,\mathcal{M}^\dagger\stackrel{!}{=}\mathcal{X}\;.
\end{equation}
Depending on whether we assume $\mathcal{M}$ to be a flavor- or CP-type transformation, see Eq.~\eqref{eq:generalM},
this demands either
\begin{equation}\label{eq:Aut_CP1}
AA^\mathrm{T} = A^* A^\dagger =\mathbbm{1}_2\;, \qquad \mathrm{or} \qquad BB^\mathrm{T} = B^* B^\dagger =\mathbbm{1}_2\;.
\end{equation}
Therefore, $A$ and $B$ have to be real orthogonal matrices $A,B\in \mathrm{O}(2)$. 

As a remark, note that the branching of the CP1-generating $\mathcal{X}$ into (\Z2) eigenstates is 
$\mathbbm{1}_0\oplus\mathbbm{1}_0\oplus\mathbbm{1}'\oplus\mathbbm{1}'$ corresponding to the CP-eigenstates $\hat\Phi_{a,\pm}:=\frac{1}{\sqrt{2}}\left(\Phi_a\pm\Phi_a^*\right)$.
The residual $\mathrm{O}(2)$ freedom we are discussing here is precisely the remaining freedom of rotating or reflecting these eigenstates 
in the space $a=1,2$.

\subsubsection{Flavor-type Outs of CP1}
Let us first focus on flavor-type Outs of CP1, i.e.\ transformations generated by $\mathcal{M}_{\mathrm{flav}}$ with $A\in \mathrm{O}(2)$. 
In general, $\mathrm{O}(2)$ can be split into reflections and rotations with explicit realizations
\begin{equation}
\label{eq:CP1-autsA}
      A_{\mathrm{ref}}(\theta)~=~\mat{\cos\theta & \sin\theta \\ \sin\theta & -\cos\theta},
      \qquad\text{and}\qquad
      A_\mathrm{rot}(\theta)~=~\mat{\cos\theta & -\sin\theta\\ \sin\theta & \cos\theta},      
\end{equation}
of determinant $-1$ and $1$, respectively. The reflections generate \Z2 inversions, 
while the rotations generate Abelian rotations from \Z{2,3,\dots} up to $\SO2\cong\U1$.
Which of these symmetry groups are realizable
by the Higgs potential cannot be decided by inspecting the symmetry transformations, but instead, crucially depends on the specific present couplings
of the potential and the order of truncation in the sense of an effective field theory.\footnote{%
Allowing for additional invariant, effective operators beyond dimension $4$ renders the theory power-counting non-renormalizable but enhances the possibilities for realizable symmetry groups.
Note that this introduces additional covariantly transforming objects under general basis changes, which supports our view that determining the possible realizable symmetry groups directly from 
the possible alignments of covariants makes sense. We note that the larger accidental symmetry group would still be realized in the far IR when irrelevant operators stop mattering due to the RG flow.}
We will focus entirely on the power-counting renormalizable potential here and, therefore, 
only consider operators up to mass dimension four,
as shown in Eq.~\eqref{eq:Potential}. To determine
which symmetry generators are ultimately conserved, it is convenient to check the basis invariant conditions for the respective symmetry groups formulated in~\cite{Bento:2020jei}.

\paragraph{\Z2 as an Out of CP1.}
The \Z2 reflections generated by $\mathcal{M}_{\mathrm{flav}}$ with $A_\mathrm{ref}(\theta)$ corresponds to the usual \Z2 class of symmetries of the 2HDM.
The easiest way to see this is to note that CP1 transformations are invariant under purely orthogonal basis transformations. 
Hence, such transformations may be used to rotate $A_{\mathrm{ref}}(\theta)$ into a more common form. 
For example, it can be transformed the most commonly considered \Z2 generators, namely the Pauli matrices $\sigma_1$ or $\sigma_3$
corresponding to choices of $\theta=\pi/2$ or $\theta=0$, respectively. 
This shows that the \Z2 class of symmetries is obtained from the flavor-type reflection Outs of CP1.

Let us label the corresponding \Z2's as $\Z{2,i=1,2,3}$ corresponding to the respective Pauli matrix which generates them.
Note that neither $\Z{2,1}$ nor $\Z{2,3}$ as symmetry generators imply the conservation of CP1 in the form of $\mathcal{X}$ in Eq.\,\eqref{eq:CP1} above.
Nonetheless, imposing any individual \Z2 always does imply the conservation of a symmetry of the CP1 class as an accidental symmetry of the potential.
One can use dedicated basis rotations to show that these also imply CP1. Hence, imposing any \Z2 generated by $A_{\mathrm{ref}}(\theta)$ together with CP1 yields the \Z2 category of 2HDM symmetries. 
In the notation adopted from \cite{Ferreira:2009wh} this reads $\mathrm{CP}1\oplus\Z2=\Z2$.

As a side remark, note that it is not possible to obtain $\Z{2,2}$ from $A_{\mathrm{ref}}(\theta)$ by a purely orthogonal basis rotation. 
This means that the flavor-type transformation generated by $\Z{2,2}$ is not contained in the normal extensions of CP1.
Requesting it nonetheless as a preserved transformation corresponds to an unorthodox extension of CP1 which should yield a different group. 
Indeed, requiring $\Z{2,2}$ together with the specific form of CP1 in Eq.\,\eqref{eq:CP1}
does not correspond to the \Z2 class of symmetries but to the CP2 class. 

\paragraph{CP2 from a flavor-type Out of CP1.}
We next discuss the rotations with $A_\mathrm{rot}(\theta)$. By themselves, these rotations generate an $\SO2\equiv\U1_2$, where again we 
label different possible \U1's as $\U1_i$ where $i=1,2,3$ corresponds to the respective Pauli matrix whose exponentiation generates 
the \U1. Since \SO2 is abelian, $A_\mathrm{rot}(\theta)$ is insensitive to orthogonal basis transformations. 
Let us first restrict the rotation to specific angles in order to obtain subgroups of $\SO2$.
Taking $\theta=\pi/2$ one obtains a \Z2 rotation that, together with CP1 implies the CP2 symmetry of the
potential. Taking other fractions of $2\pi$, such as for example $2\pi/3$, which would lead to a \Z3 rotation, 
directly leads to the full \SO2 symmetry. This is, as is well known, simply due to the fact that the potential 
(truncated at the renormalizable order) cannot realize the \Z3 and higher-order symmetries.

\paragraph{CP3 from a flavor-type Out of CP1.}
Imposing the full $\SO2\cong\U1_2$ generated by $A_\mathrm{rot}(\theta)$ as additional preserved transformation certainly implies that we are \textit{at least} in the \U1 class of 2HDM symmetries. However, the additional requirement of CP1 here implies that the resulting symmetry class is, in fact, CP3 and therefore larger than \U1.
Using the notation adopted from \cite{Ferreira:2009wh} this would read $\mathrm{CP}1\oplus\U1_2=\mathrm{CP}3$.
This equivalence is straightforward to see in the $\rep{r}\oplus\rep{r}^*$ space noticing that $(i)$ the CP3 generator is just the direct combination of the CP1 and $\U1_2$ 
generators, and $(ii)$ that $\U1_2$ is generated by a squared CP3 transformation, while CP1 is included in the CP3 transformation for the special value~$\theta=0$.

\paragraph{U(1) from unorthodox extension of CP1, or from flavor-type Out of \Z2.}
The above discussion shows that it is not straightforward to obtain the \U1 class of symmetries from Outs of CP1 
because the corresponding normal extension by $\SO2=\U1_2$ would enlarge the group to CP3. Also, it is clear that $\U1_{1}$ or $\U1_{3}$, 
which would realize the \U1 class of symmetries on top of CP1, are not included as solutions of the consistency condition, Eq.~\eqref{eq:ccCP1},
because they do not correspond to orthogonal matrices, hence, violate the constraint imposed by Eq.~\eqref{eq:Aut_CP1}.
This shows that CP1 is not normal with respect to $\mathrm{CP1}\oplus\U1_{1,3}$, implying that the corresponding extension of CP1 to $\U1_{1,3}$ is of the unorthodox type.

Nonetheless, we stress that it is possible to obtain $\U1$ from a normal group extension if we do not start from CP1 but 
directly from its (normal) extension \Z2. Without loss of generality, let us take the extension of CP1 generated by $\Z{2,3}$.
Starting from $\Z{2,3}$, this group can be normally extended by a trivial flavor-type Out to $\U1_3$ (which is easy to see 
because a matrix commutes with its matrix exponential). We note that neither of these transformations will generate CP1, but both of them will imply an accidental CP1 in the renormalizable 2HDM potential.

\subsubsection{CP-type Outs of CP1}
Next, we consider the CP-type Outs of CP1, that is transformations generated by $\mathcal{M}_{\mathrm{CP}}$ with $B\in \mathrm{O}(2)$. 
Again, $\mathrm{O}(2)$ can be split into reflections and rotations with explicit realizations
\begin{equation}
\label{eq:CP1-autsB}
      B_{\mathrm{ref}}(\theta)~=~\mat{\cos\theta & \sin\theta \\ \sin\theta & -\cos\theta},
      \qquad\text{and}\qquad
      B_\mathrm{rot}(\theta)~=~\mat{\cos\theta & -\sin\theta\\ \sin\theta & \cos\theta}.    
\end{equation}
The reflection transformation can again be rotated by an orthogonal basis change to any of the Pauli matrices $\sigma_{1,3}$ without affecting the CP1 transformation. 
This shows that $\mathrm{CP1}\oplus B_{\text{ref}}(\theta)=\Z2$. Alternatively, note that this is easy to see because the  two generators together, taken as $4\times4$ matrices 
in the space $\rep{r}\oplus\rep{r}^*$ generate exactly the same matrix group as $\mathrm{CP1}\oplus A_{\mathrm{ref}}(\theta)$ discussed above.

Lastly, we discuss the CP-type transformations with $B_\mathrm{rot}(\theta)$. It is easy to see that for discrete rotations of an angle $\theta=\pi/2$
those generate CP2, while for all smaller angles or even continuous choices of $\theta$ this generates CP3.
Both of these symmetries are realizable even if CP1 is required in addition.

\subsection{SU(2) as an Out of CP2}
General \SU{2} transformations can never arise as a flavor or CP-type outer automorphism of CP1 because the corresponding consistency condition, Eq.~\eqref{eq:ccCP1}, leads to the constraint \eqref{eq:Aut_CP1} which requires orthogonal matrices. In fact, \SU{2} transformations generate unorthodox extensions of CP1, and we will come back to this below.

Let us first show that \SU{2} transformations do arise as outer automorphisms, i.e.\ normal extensions, of CP2. For a CP2 transformation, $\mathcal{X}$ takes the form
\begin{equation}
 \mat{\Phi^{\hphantom{*}}\\ \Phi^*}\xmapsto{\mathrm{CP2}} \mathcal{X}_\mathrm{CP2} \mat{\Phi^{\hphantom{*}}\\ \Phi^*} = \mat{  0&  \varepsilon \\   \varepsilon &  0}\mat{\Phi^{\hphantom{*}}\\ \Phi^*}\;,
\end{equation}
with the totally antisymmetric tensor $\varepsilon$. 
The corresponding consistency condition for a trivial outer automorphism is the same as Eq.~\eqref{eq:ccCP1} replacing $\mathcal{X}$ by $\mathcal{X}_\mathrm{CP2}$. For flavor- or CP-type transformation, respectively, $A$ and $B$ in $\mathcal{M}$ are constrained to fulfill 
\begin{equation}\label{eq:Aut_CP2}
A\varepsilon A^\mathrm{T} = A^*\varepsilon A^\dagger =\varepsilon\; \qquad \mathrm{or} \qquad B\varepsilon B^\mathrm{T} = B^*\varepsilon B^\dagger =\varepsilon\;.
\end{equation}
These automatically hold for all elements of \SU{2} simply because $\varepsilon$ is  an invariant tensor of this group. 
Irrespectively of whether we extend CP2 by a flavor- or CP-type \SU{2} transformation the result is the same and yields the full \SU{2} symmetry of the 2HDM potential.
This discussion also shows that we can normally extend CP2 to CP3 by either a flavor- or CP-type transformation if the Out transformation is restricted to $A$, $B\in\SO{2}$. 

\subsection{SU(2) as an unorthodox extension of CP1}
Finally, we note that \SU{2} symmetry can also be obtained from an extension of the CP1 transformation $\Z2^\mathrm{CP1}$ generated by $\mathcal{X}_\mathrm{CP1}$. However, the corresponding extension is unorthodox, see Sec.~\ref{sec:extensions}. 
For this, note that
\begin{equation}
 \mathcal{M}\,\mathcal{X}_{\mathrm{CP1}}\,\mathcal{M}^\dagger~=~\mathcal{X}_{\mathrm{CP1}}'
\end{equation}
enforces a $\Z2$ CP-type transformation $\mathcal{X}_{\mathrm{CP1}}'$ which is generally distinct from $\Z2^\mathrm{CP1}$ for all possible non-orthogonal flavor or CP-type transformations $\mathcal{M}$ with $A\in\SU{2}$ or $B\in\SU{2}$. There is an infinite family of correspondingly conjugate subgroups of $\Z2^\mathrm{CP1}$ in the big group $\Gamma$, which is then of course enlarged to the full $\SU{2}\cup\mathrm{CP1}$. 

We remark that the generators of \SU{2} do not by themselves generate CP1 (or any other CP-type transformation). Hence, it is again an accident of the renormalizable 2HDM potential that requiring \SU{2} from the get-go also implies CP1 (and, therefore, from $\SU{2}\cup\mathrm{CP1}$, also CP2 and CP3).

\subsection{Remarks}\label{sec:remarks}
We have shown that we can generate the \Z2, CP2, and CP3 class of symmetries of the 2HDM as Outs of CP1. 
Furthermore, \U1 can be generated as Out of \Z2, and CP3 as well as \SU{2} can be generated as Outs of CP2.
This explicitly demonstrates that we can reach all realizable symmetries of the 2HDM by extending the smallest possible transformation CP1 by consecutive outer automorphisms.
Additionally, we have shown that CP3 can be generated from an unorthodox extension of \U1, and \SU{2} can be generated from an unorthodox extension of CP3. It is, furthermore, straightforward to show that CP2 corresponds to an unorthodox extension of \Z{2} (even though the corresponding transformation does, in fact, close up to a global hypercharge rotation).
This shows how every arrow in Fig.\,\ref{fig:SymmetryMap} can be understood as a group extension, and clarifies the normal or unorthodox nature of the extensions. 

The fact that all possible symmetry groups of the 2HDM can be reached by Outs is non-trivial. There is no general reason why this should be the case in a generic model. 
In fact, there is a simple argument which shows that this is not the case in general: By construction, Outs (as normal extensions) imply that the small group $G$ is a normal subgroup of the big group $\Gamma$. However, there is the class of simple groups which are defined as to have no non-trivial normal subgroups. Hence, simple groups can never be reached by a normal group extension. Therefore, any individual model which allows for a realizable simple group acts as an example that not all symmetry groups can be obtained through normal extensions. 
In such a case, unorthodox extensions are required in order to obtain all the symmetries of the model by extension of smaller groups. 

\section{Symmetries of the 3HDM from outer automorphisms}
To provide a second example, we briefly demonstrate that the principle outlined above also works for the 3HDM.
We focus on the finite realizable Higgs-flavor type of symmetries\footnote{%
The discussion for CP-type and/or continuous transformations works completely analogously. For their classifications see, respectively, \cite{Ivanov:2012fp,Ivanov:2014doa} and \cite{deMedeirosVarzielas:2019rrp,Darvishi:2019dbh,Kuncinas:2024zjq}.}
shown in the ``symmetry tree'' of \cite[Fig.~1]{Ivanov:2014doa} (see also \cite{Ivanov:2012fp}) and reproduced in slightly modified form in our Fig.~\ref{fig:ModifiedTree}.
The l.h.s.\ of Fig.~\ref{fig:ModifiedTree} shows the names of the corresponding groups. Arrows denote that a group $G$ is a subgroup of the group $\Gamma\supset G$ to which the arrow is pointing to. The r.h.s.\ of Fig.~\ref{fig:ModifiedTree} shows possible explicit choices of generators of the respective groups in their three-dimensional representation acting on the three Higgs fields. An explicit choice for the matrix generators is~\cite{Ivanov:2014doa} (note that we use a slightly modified form for $a_4$)
\begin{align}\notag
\sigma_{12} &:=
\begin{pmatrix}
-1 & 0 & 0\\
0 & -1 & 0\\
0 & 0 & 1
\end{pmatrix},
\quad
\sigma_{23}:=
\begin{pmatrix}
1 & 0 & 0\\
0 & -1 & 0\\
0 & 0 & -1
\end{pmatrix},
\quad
a_3:=
\begin{pmatrix}
1 & 0 & 0\\
0 & \omega & 0\\
0 & 0 & \omega^2 
\end{pmatrix},\quad
a_4:=
\begin{pmatrix}
1 & 0 & 0\\
0 & 0 & -1\\
0 & 1 & 0
\end{pmatrix}\,,
&
\\\label{eq:3HDM_generators}
c&:=-
\begin{pmatrix}
1 & 0 & 0\\
0 & 0 & 1\\
0 & 1 & 0
\end{pmatrix},
\quad
b:=
\begin{pmatrix}
0 & 1 & 0\\
0 & 0 & 1\\
1 & 0 & 0
\end{pmatrix},
\quad
d:=\frac{i}{\sqrt{3}}
\begin{pmatrix}
1 & 1 & 1\\
1 & \omega^2 & \omega\\
1 & \omega & \omega^2
\end{pmatrix},\quad
\text{with}\quad \omega:=e^{\frac{2\pi i}{3}}\,.
& \raisetag{30pt}
\end{align}
This choice of generators is not unique. Even though the generators stated in Fig.~\ref{fig:ModifiedTree} do not always form minimal generating sets we present the groups in this way for convenience of the discussion. Also, it is sometimes necessary to perform a basis change on the starting group before some of the subgroup relationships in Fig.~\ref{fig:ModifiedTree} become manifest and we will come back to this below.

We distinguish multiple types of arrows (solid thick, solid, dashed, dotted) in Fig.~\ref{fig:ModifiedTree}. We first discuss the solid arrows. Solid arrows denote normal group extensions from $G$ to $\Gamma$, i.e.\ $G\triangleleft\Gamma$. This implies that any $\mathsf{h}\in\Gamma$ solves
the consistency condition Eq.~\eqref{eq:group_extension}, i.e.\ $\mathsf{h}\,G\,\mathsf{h}^{-1}=G\;\forall\mathsf{h}\in\Gamma$ (it is enough to verify this for the generators of $\Gamma$ and $G$). 
For example, consider the case $G=\mathbbm{Z}_4=\langle a_4 \rangle$ and $\Gamma=D_4=\langle a_4,c\rangle$. It is easy to confirm that
\begin{align}
&\Z{4}\rightarrow D_4:&
&c\,a_4\,c^{-1}~=~a_4^{-1}\;,& &\text{and, trivially,}& &a_4\,a_4\,a_4^{-1}~=~a_4\;.&
\end{align}
This shows that $\mathbbm{Z}_4\triangleleft D_4$, and that $D_4\cong\mathbbm{Z}_4\rtimes\mathbbm{Z}_2$ is the split extension of $\mathbbm{Z}_4$ by its $\mathbbm{Z}_2$ outer automorphism, here generated by $c$. 
Analogous considerations can be made for all the solid arrows in Fig.~\ref{fig:ModifiedTree}:\enlargethispage{1cm}%
\tikzset{node distance=1.2cm, auto}
\begin{figure}[!t!]
\begin{minipage}[l]{0.5\linewidth}
\scalebox{0.85}{
\begin{tikzpicture}
  \node (Z2) {$\mathbbm{Z}_2$};
  \node (Z2Z2) [node distance=1.0cm,above of=Z2] {$\mathbbm{Z}_2\times\mathbbm{Z}_2$};
  \node (Z3) [node distance=2.5cm, right of=Z2Z2] {$\mathbbm{Z}_3$};
  \node (Z4) [node distance=2.5cm, left of=Z2Z2] {$\mathbbm{Z}_4$};
  \node (D8) [above of=Z4] {$D_4$};
  \node (A4) [above of=Z2Z2] {\phantom{$A_4$}};
  \node (S4) [node distance = 1.4cm, above of=A4] {$S_4$};
  \node (D6)[node distance=2.5cm, right of=A4]{$S_3$};
  \node (Z3Z3) [node distance=2.0cm, right of=Z3] {$\underline{\mathbbm{Z}_3\times\mathbbm{Z}_3}$};
  \node (D54) [above of=Z3Z3] {$\Delta(54)/\mathbbm{Z}_3$};
  \node (S36) [node distance=2.7cm,above of=D54] {$\Sigma(36)$};
  \draw[->] (Z2) to node {} (Z4);
  \draw[->] (Z2) to node {} (Z2Z2);
  \draw[dashed,->] (Z2) to node {} (D6);
  \draw[very thick,->] (Z3) to node {} (D6);
  \draw[dashed,->] (Z3) to node {} (A4);
  \draw[very thick,->] (Z4) to node {} (D8);
  \draw[very thick,->] (Z2Z2) to node {} (D8);
  \draw[very thick,->] (Z2Z2) to node {} (A4);
  \draw[very thick,->] (A4) to node {} (S4);
  \draw[dashed,->] (D8) to node {} (S4);
  \draw[dashed,->] (D6) to node {} (S4);
  \draw[dashed,->] (D6) to node {} (D54);
  \draw[very thick,->] (Z3Z3) to node {} (D54);
  \draw[->] (Z3) to node {} (Z3Z3);
  \draw[very thick,->] (D54) to node {} (S36);
  \draw[dashed,->] (Z4) to node {} (S36);
  \node (A4) [above of=Z2Z2] {$A_4$};
\end{tikzpicture}}
\end{minipage}%
\begin{minipage}[l]{0.5\linewidth}
\scalebox{0.85}{
 \begin{tikzpicture}
	\node (Z2) {$\langle\sigma_{23}\rangle$};
	\node (Z2Z2) [node distance=1.0cm,above of=Z2] {$\langle\sigma_{23},\sigma_{12}\rangle$};
	\node (Z3) [node distance=2.5cm, right of=Z2Z2] {$\langle b\rangle$};
	\node (Z4) [node distance=2.5cm, left of=Z2Z2] {$\langle a_4\rangle$};
	\node (D8) [above of=Z4] {$\langle a_4,c\rangle$};
	\node (A4) [above of=Z2Z2] {\phantom{$\langle\sigma_{23},\sigma_{12},b\rangle$}};
	\node (S4) [node distance = 1.4cm, above of=A4] {$\langle a_4,\sigma_{23},\sigma_{12},b,c\rangle$};
	\node (D6)[node distance=2.5cm, right of=A4]{$\langle b,c\rangle$};
	\node (Z3Z3) [node distance=2.0cm, right of=Z3] {$\langle b,\omega\mathbbm{1}\rangle$};
	\node (D54) [above of=Z3Z3] {$\langle a_3,c,b\rangle$};
	\node (S36) [node distance=2.7cm,above of=D54] {$\langle a_3,c,b,d\rangle$};
	\draw[->] (Z2) to node {} (Z4);
	\draw[->] (Z2) to node {} (Z2Z2);
	\draw[dotted,thick,->] (Z2) to node {} (D6);
	\draw[very thick,->] (Z3) to node {} (D6);
	\draw[dashed,->] (Z3) to node {} (A4);
	\draw[very thick,->] (Z4) to node {} (D8);
	\draw[very thick,->] (Z2Z2) to node {} (D8);
	\draw[very thick,->] (Z2Z2) to node {} (A4);
	\draw[very thick,->] (A4) to node {} (S4);
	\draw[dashed,->] (D8) to node {} (S4);
	\draw[dashed,->] (D6) to node {} (S4);
	\draw[dashed,->] (D6) to node {} (D54);
	\draw[very thick,->] (Z3Z3) to node {} (D54);
	\draw[->] (Z3) to node {} (Z3Z3);
	\draw[very thick,->] (D54) to node {} (S36);
	\draw[dotted,thick,->] (Z4) to node {} (S36);
	\node (A4) [above of=Z2Z2] {$\langle\sigma_{23},\sigma_{12},b\rangle$};
	\end{tikzpicture}
	}
\end{minipage}
\caption{(``Igor's tree'')
\label{fig:ModifiedTree}
The flavor-type finite realizable symmetry groups of the 3HDM, adopted from~\cite{Ivanov:2012fp,Ivanov:2014doa} and modified for our purpose. 
All arrows denote subgroup relationships. Solid arrows indicate
normal extension (outer automorphism), with thick(thin) solid arrows corresponding to non-trivial(trivial) outer automorphisms. Dashed arrows indicate unorthodox extensions (see classification in Sec.~\ref{sec:extensions_generalities}). The underlined Abelian group $\mathbbm{Z}_3\times\mathbbm{Z}_3$ is not realizable but included to explicitly show that all groups can be obtained through normal extensions. On the r.h.s.\ we show possible choices of explicit matrix generators of the respective groups, see Eq.~\eqref{eq:3HDM_generators}. 
Dotted arrows here indicate that the explicitly stated generators need to be basis-transformed to make the subgroup relation manifest. 
}
\end{figure}
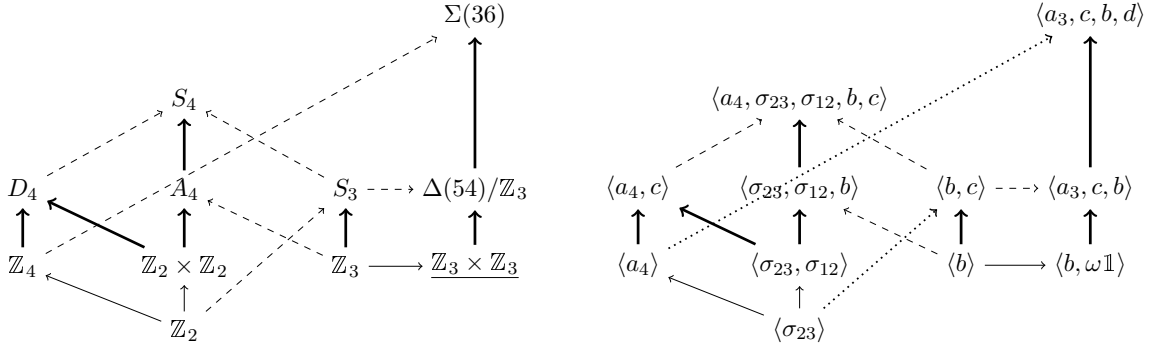%
\begin{align}\notag
&\Z{2}\rightarrow\Z{4}:&
&a_4\,\sigma_{23}\,a_4^{-1}=\sigma_{23}\,, \hspace{3.1cm} \text{(non-split $[a_4^2=\sigma_{23}]$, trivial Out)}\;,& \\\notag
&\Z{2}\rightarrow\Z{2}\times\Z{2}:&
&\sigma_{12}\,\sigma_{23}\,\sigma_{12}^{-1}=\sigma_{23}\;,\hspace{5.35cm}\text{(split, trivial Out)}\;,& \\\notag
&\Z{2}\times\Z{2}\rightarrow D_4:&
&c\,\sigma_{23}\,c^{-1}=\sigma_{23}\,,~c\,\sigma_{12}\,c^{-1}=\sigma_{12}\sigma_{23}\equiv\sigma_{13}\,, \hspace{1.0cm}\text{($\Z{2}\subset\mathrm{Out}(\Z{2}\times\Z{2}))\,,$}&\\\notag
&\Z{2}\times\Z{2}\rightarrow A_4:&
&b\,\sigma_{12}\,b^{-1}=\sigma_{13},~b\,\sigma_{13}\,b^{-1}=\sigma_{23},~b\,\sigma_{23}\,b^{-1}=\sigma_{12},\text{($\Z{3}\subset\mathrm{Out}(\Z{2}\times\Z{2}))\,,$}&\\\notag
&A_4\rightarrow S_4:& 
&c\,b\,c^{-1}=b^{-1}\,,~c\,\sigma_{23}\,c^{-1}=\sigma_{23}\,,~
c\,\sigma_{12}\,c^{-1}=\sigma_{13}\,,~\hspace{0.85cm}\text{($\Z{2}\cong\mathrm{Out}(A_4))\;,$}& \\\notag
&\Z{3}\rightarrow S_3:& 
&c\,b\,c^{-1}=b^{-1}\,,\quad\hspace{6.3cm}\text{($\Z{2}\cong\mathrm{Out}(\Z{3})$)\,,}&
\\\notag
&\Z{3}\rightarrow \Z{3}\times\Z{3}:& 
&e_3\,b\,e_3^{-1}=b\,\quad(e_3:=\omega\mathbbm{1})\quad\hspace{4.1cm}\text{(split, trivial Out)\,,}& \\\notag
&\Z{3}\times\Z{3}\rightarrow \Delta(54):& 
&a_3\,b\,a_3^{-1}=e_3\,b\,,~a_3\,e_3\,a_3^{-1}=e_3\,,~c\,b\,c^{-1}=b^{-1}\,,~c\,e_3\,c^{-1}=e_{3} \\[-0.15cm]\notag
&& &\omit\hfill \text{(split, $S_3^{(a_3,c)}\subset\mathrm{Out}(\Z{3}\times\Z{3})$)\,,} & \\\notag
&\Delta(54)\rightarrow \mathrm{SG}(108,15):& 
&d\,c\,d^{-1}=c\,,~d\,b\,d^{-1}=a_3\,,~d\,a_3\,d^{-1}=b^2\,,& \\[-0.15cm]\notag
&&
&\omit\hfill\text{(non-split $[d^2=c]$, $\Z{4}\subset\mathrm{Out}(\Delta(54))\,.$}&
\end{align}
In this way all possible realizable finite groups can be reached by consecutive normal extensions of the smallest ``seed'' symmetry groups by outer automorphisms. Thin arrows in Fig.~\ref{fig:ModifiedTree} denote trivial outer automorphisms. We note that reaching $\Delta(54)/\mathbbm{Z}_3$ via a normal extension of a subgroup requires to include the non-realizable group $\mathbbm{Z}_3\times\mathbbm{Z}_3$ (underlined in Fig.~\ref{fig:ModifiedTree}), which itself is a normal extension of a realizable $\mathbbm{Z}_3$ symmetry. We also remark that $\langle a_3,b,c\rangle=\Delta(54)$, $\langle a_3,b,c,d\rangle=\mathrm{SG}(108,15)$ (this group is often also called $\Sigma(36\times3)$, see~\cite{Hagedorn:2013nra}) while the groups in Fig.~\ref{fig:ModifiedTree} are denoted as $\Delta(54)/\mathbbm{Z}_3\cong\mathrm{SG}(18,4)$ and $\Sigma(36)\cong\mathrm{SG}(36,9)\cong\mathrm{SG}(108,15)/\mathbbm{Z}_3$.\footnote{%
``$\mathrm{SG}$'' here denotes the \texttt{GAP} \texttt{SmallGroup} label of the respective groups~\cite{GAP4}.}
This is because their non-trivial $\mathbbm{Z}_3$ center can be absorbed by a global hypercharge rotation.

Now consider the dashed arrows in Fig.~\ref{fig:ModifiedTree}. Those correspond to non-normal subgroup relations. Hence, advancing from a group $G\subset\Gamma$ to $\Gamma$ via a dashed arrow requires an unorthodox extension. 
This implies that there is an element $\mathsf{h}\in\Gamma$ which does not map $G$ onto itself, but to a conjugate isomorphic subgroup $\mathsf{h}\,G\,\mathsf{h}^{-1}=G'\neq G$. Consider, for example, the connection from $\mathbbm{Z}_3=\langle b\rangle$ to $A_4=\langle b, \sigma_{12}\rangle$.
Since $\mathbbm{Z}_3$ is not normal in $A_4$ a normal group extension is not possible. Instead, extending $\langle b\rangle$ by $\sigma_{12}$ yields
\begin{align}
 &\mathbbm{Z}_3\rightarrow A_4:& &\sigma_{12}\,b\,\sigma_{12}^{-1}~=~b'\;,\qquad\text{where}\qquad \langle b'\rangle=\mathbbm{Z}_3^{\prime}\neq\langle b\rangle\;.&
\end{align}
$b'=b^2\,\sigma_{12}\,b^2=\sigma_{23}\,b=b\,\sigma_{13}$ here generates a group that is isomorphic but not identical to the group generated by~$b$. 
In the same manner one can find unorthodox extensions along all the dashed arrows in Fig.~\ref{fig:ModifiedTree}:
\begin{align}\notag
&D_4\rightarrow S_4:&
&b\,c\,b^{-1}=b^2\,c\,,~b\,a_4\,b^{-1}=b^2\,a_4^{-1}\,,& &\text{(unorthodox)}\;,& \\
&S_3\rightarrow S_4:&
&\sigma_{12}\,b\,\sigma_{12}^{-1}=b^2\,\sigma_{12}\,b^2,~\sigma_{12}\,c\,\sigma_{12}^{-1}=c\,\sigma_{12}\,c\,\sigma_{12}\,c\,,& &\text{(unorthodox)}\;,& \\\notag
&S_3\rightarrow \Delta(54):&
&a_3\,b\,a_3^{-1}=\left(b\,a_3\,b\right)^{-1}a_3,~a_3\,c\,a_3^{-1}=a_3^2\,c\,,& &\text{(unorthodox)}\;.&
\end{align}
We note that the relations always have to close within the two generators on the l.h.s.\ of each equation, and that the r.h.s.\ can always be written as a translation (by the commutator element) of the respective seed group generator.

Finally, we discuss the dotted arrows in Fig.~\ref{fig:ModifiedTree}. Note that with the explicit choice of generators, indicated on the r.h.s.\ of Fig.~\ref{fig:ModifiedTree} and in Eq.~\eqref{eq:3HDM_generators}, it is necessary to switch basis for the seed generators before a subgroup relation can be established along the dotted arrows. For example, $\Z{4}=\langle a_4\rangle$ is not a subgroup of $\mathrm{SG}(108,15)=\langle b,d\rangle$. However, the whole tree can also be constructed starting from the seeds $\Z2=\langle c\rangle$, $\Z{4}=\langle d\rangle$ in which case the subgroup relation $\Z{4}^{\langle d \rangle}\subset\langle b,d\rangle$ is manifest.
It is not possible to chose a basis for all the seed generators in Fig.~\ref{fig:ModifiedTree} to simultaneously realize all subgroup relations. There is a simple argument that proves this: Consider the lines from $\Z{2}\rightarrow S_3$ and from $\Z{3}\rightarrow A_4$.
If $\langle\Z2,\Z3\rangle=S_3$, these very same $\Z2$ and $\Z3$ generators cannot be subgroups of $A_4$, since $A_4$ does not have a $S_3$ subgroup. Hence, $A_4$ and $S_3$ must recruit their seeds from at least partially distinct generators which proves our point. We note, however, that it is possible to chose a basis for the whole tree such that all ``additionally needed'' generators can be recruited from the (computable) outer automorphisms of the groups.

We conclude that also for the 3HDM it is, in principle, possible to find all of the realizable finite symmetry groups by consecutively exploiting normal extensions by outer automorphisms starting from the smallest possible symmetry groups. Of course, we have not \textit{derived} all the finite realizable symmetry groups of the 3HDM here, but just have shown that the previously obtained results in the literature are consistent with our logic of using redundancies of present transformations in order to identify possible extensions. We also want to stress explicitly that it is no surprise that all realizable finite symmetry groups of the 3HDM can be reached by normal group extensions, as this was the way that they were originally derived in~\cite{Ivanov:2012fp,Ivanov:2014doa} (without the notion of outer automorphisms). Nonetheless, it is known that normal extensions do not cover all possible extensions in more general models\footnote{%
For example, for NHDM with $N\geq4$ Burnside's $pq$-theorem is not enough to guarantee that all arising groups are solvable~\cite{Shao:2023oxt,Shao:2024ibu}. See also the discussion about realizable simple groups in Sec.~\ref{sec:remarks}.}, where dealing with unorthodox extensions becomes a necessity. Our treatment, hence, adds to the existing literature because we introduce the use of outer automorphisms, include unorthodox extensions, and provide a treatment that seamlessly includes (normal and unorthodox) extensions also by continuous groups.

\section{Conclusions}
We have discussed extensions of symmetry groups of generic QFTs in such a way that the enhanced symmetry contains the starting seed symmetry group as a subgroup.
We have shown that such extensions are given either by so-called normal extensions, which are realized by outer automorphisms, or, by unorthodox extensions that we have introduced here (see Fig.~\ref{fig:extensions}). The 2HDM and 3HDM are used as pedagogical examples of our general discussion of symmetry extensions, since the global symmetry groups of their potentials are known and have previously been classified.

We have reiterated the fact that the action of symmetries, in general, can have two different kinds of physical consequences on the
co- and invariant combinations (under basis changes) of couplings of a theory, and how this fact can be exploited to construct a 
two-dimensional symmetry map for the 2HDM, Fig.~\ref{fig:SymmetryMap}. We have then shown how all symmetries of the 2HDM, including continuous and discrete, flavor- and CP-type transformations, 
can be obtained via consecutive extension by outer automorphisms of the smallest possible seed symmetry, CP1. 
We note that this introduces the first example of a continuous outer automorphism in the physics literature for some time.
We have also shown how some of the direct steps in the 
symmetry map of the 2HDM can only be understood as unorthodox extensions. 

As a second example, we have discussed the discrete flavor-type symmetry groups of the 3HDM. Also in this case, all of the groups can be obtained by consecutively extending the possible smallest seed symmetry groups by outer automorphisms. Here, this did not come as a surprise, as the whole tree of these groups, Fig.~\ref{fig:ModifiedTree}, was originally obtained using normal group extensions~\cite{Ivanov:2012ry,Ivanov:2012fp,Ivanov:2014doa}.
Also the feature that some symmetry enhancements can be obtained from outer automorphism was known, specifically for the enhancement from $\Delta(54)/\Z{3}\rightarrow\Sigma(36)$~\cite{Fallbacher:2015rea}.
Also for the 3HDM, we have shown that some of the steps in the symmetry tree can only be understood as unorthodox extension.

One may have the impression that our discussion is not terribly useful for the 2HDM/3HDM, where all possible symmetry groups and a large part of the extensions have been known before.
What is new here is that we clearly point out that all normal extensions are given by outer automorphisms (including the trivial one), and that this also applies to CP-type and continuous symmetry transformations. 
Furthermore, we introduced the notion of unorthodox extensions and emphasized the fact that these are, in general, required to exist specifically in models with realizable simple groups. Finally, our insights are extremely useful for models with so far unexplored, potentially large symmetry groups and landscapes. Being able to compute the allowed extensions (in the case of non-trivial outer automorphisms), or at least being able to constrain the extensions (in the case of unorthodox extensions) decisively limits the space of possible extensions. This can meaningfully improve brute force scans and inform machine learning algorithms. We also stress that the knowledge about the possible symmetry extensions allows to compute the fixed boundaries of the RGE flow where the seed symmetry is enhanced by the respective extension~\cite{Trautner:2016ezn}.

Finally, it is, in general, not trivial to decide whether a given symmetry group is realizable for a given QFT. To decide this in a ``bottom-up'' way usually requires a good knowledge about other possible symmetries of the theory and their explicit action on the couplings. We have suggested a way how this obstacle can be overcome in the future by using the covariantly transforming irreducible basis-covariant 
combinations of couplings and their possible relative alignments, which is expected to give a ``top-down'' way to directly determine realizable symmetry groups without having to scan over all their non-realizable subgroups.

 \section*{Acknowledgements}
We thank Ingolf Bischer for participation in the early stages of this work.
We are grateful to Igor P.\ Ivanov for very useful correspondence and insightful comments on the manuscript.
The work of CD is supported by the F.R.S./FNRS under the Excellence of Science (EoS) project No.\ 30820817 - be.h ``The H boson gateway to
physics beyond the Standard Model'' and by the IISN convention No.\ 4.4503.15.

\bibliography{Bibliography}
\addcontentsline{toc}{section}{Bibliography}
\bibliographystyle{utphys}
\end{document}